# Advantage on Superconductivity of Heavily Boron-Doped (111) Diamond Films

Hitoshi Umezawa[1], Tomohiro Takenouchi[1], Yoshihiko Takano[2], Kensaku Kobayashi[1],

Masanori Nagao[2], Isao Sakaguchi[2], Minoru Tachiki[2], Takeshi Hatano[2], Guofang Zhong[1]

Masashi Tachiki[2] and Hiroshi Kawarada[1]

1 School of Science and Engineering, Waseda University, 3-4-1 Okubo, Shinjyuku-ku, Tokyo

169-8555, Japan

2. National Institute for Materials Science, 1-2-1 Sengen, Tsukuba 305-0047 Japan.

**Abstract**

The superconductivity transition temperatures $T_{c(onset)}$ of 11.4 K and $T_{c(offset)}$ of 7.4 K, which are the highest in diamond at present, are realized on homoepitaxially grown (111) diamond films with a high boron doping concentration of $8.4 \times 10^{21}$ cm$^{-3}$ (4.7 at%). Tc values of (111) diamond films are more than twice as high as those of (100) films at the equivalent boron concentration. The Tc of boron-doped (111) diamond increases as the boron content increases up to the maximum incorporated concentration and is agrees with the value estimated using McMillan's equation. The advantageous Tc for (111) diamond films is due to the higher carrier concentration which exceeds its boron concentration.

Diamond is known as a band insulator with a wide band gap of 5.5 eV. However, by introducing impurities such as boron into the substitutional sites, semiconducting diamond can be obtained. Recently, Ekimov et al. have investigated that heavily boron-doped (4.9 x $10^{21}$ $cm^{-3}$, 2.8 at%) diamond synthesized by high-pressure and high-temperature (HTHP) method exhibits type II superconductivity at 2.3 K (zero resistance) [1]. For superconductivity, they use the metallic transport property of the impurity band, which is commonly observed in diamond with boron doping levels of more than $10^{20}$ $cm^{-3}$.

The deposition of heavily boron-doped diamond films by means of chemical vapor deposition (CVD) is a useful technique for improving the quality of diamond films and controlling the density of boron. Moreover, the low-temperature deposition of CVD diamond makes it possible to fabricate superconducting devices easily. We have successfully introduced high boron concentrations and a carrier density of 9.4 x $10^{20}$ $cm^{-3}$ on polycrystalline diamond films by controlling the growth orientation to (111), and have realized Tc of 4.2 K [2]. This value is higher than the Tc of 2.1 K for heavily boron-doped (1.9 x $10^{21}$ $cm^{-3}$) diamond deposited on (100) single-crystalline diamond [3]. Due to the advantages of a very high Debye temperature (1860 K), some theoretical studies on the potential of a higher superconductivity transition temperature Tc have been carried out [4-7]. In this study, we have successfully introduced high boron concentrations more than $5 \times 10^{21}$ cm$^{-3}$ into (111) and (100) single-crystalline diamonds using CVD, and have achieved a high Tc of 7.4 K. The advantages of the superconductivity transition temperature of (111) diamond will be also discussed.

Homoepitaxial growth is carried out on the HPHT synthetic type Ib (111) and (100) diamonds. The deposition conditions were 50 Torr chamber pressure and 800-900°C substrate temperature with diluted gas mixtures of methane and trimethylboron (TMB) in hydrogen. The methane concentration is 3% in hydrogen with a B/C ratio of 2000 - 12000 ppm. Films of 1 – 4 µm thickness were used for the following measurements.

To characterize the heavily boron-doped diamond films, we employed Raman spectroscopy (Renishaw inVia Reflex system, 632.8 nm), Hall effect measurement, X-ray diffraction (XRD: θ - 2θ measurement), secondary ion mass spectroscopy (SIMS), transport and the measurement of magnetization properties using a superconducting quantum interference device (SQUID) magnetometer.

Crystal growth orientation is observed by means of the θ - 2θ measurement of XRD. The (111) and (100) growths are observed on (111) and (100) deposited diamonds, respectively. The Raman shifts of the heavily boron-doped homoepitaxially grown diamond (111) and (100) films are shown in fig. 1. (111) and (100) homoepitaxial diamond films have equivalent boron concentrations of 8.4 x $10^{21}$ $cm^{-3}$ and 8.5 x $10^{21}$ $cm^{-3}$ measured by SIMS. Those boron concentrations correspond to the atomic percentages of 4.7 and 4.9% in diamond, respectively. The two Raman spectra similarly show two large wide bands at approximately 1210-1230 $cm^{-1}$ and 450-500 $cm^{-1}$, which are typically observed in heavily boron-doped ([B] > 3 x $10^{20}$ $cm^{-3}$) diamonds [8-10]. The zone-centre optical phonon line normally observed at 1332 $cm^{-1}$ is decreased and widened at this doping level. The decreased peak of the zone-centre optical phonon is observed as a shoulder of the 1210 -1230 $cm^{-1}$ band, which is similar to the Raman spectra reported in ref. [1]. The origins of the bands at 1210-1230 and 450-500 $cm^{-1}$ has been discussed in the following [9, 10]. Gonon et al. indicated that the band at 1200 $cm^{-1}$ is attributed to the maxima in one phonon density of state of diamond, and suggested that this band comes from regions that are disordered due to the heavy boron doping [9]. The band at 500 $cm^{-1}$ is explained by Bernard et al. as being the local vibration modes (LVM) of boron pairs [10]. They also point out that this band shifts to lower wavenumbers according to the density of the boron pairs in diamond. In our case, the band center of LVM from the boron pairs on (111) diamond shifts to be lower wavenumbers than that of (100) diamond, which suggests that a more dense boron pairs exists in the (111) diamond film even for equivalent boron contents (8.4 x $10^{21}$ cm$^{-3}$ for (111), 8.5 x $10^{21}$ cm$^{-3}$ for (100)), respectively. The broad band at 1490 cm$^{-1}$ on (111) diamond corresponds to high resistivity surface amorphous carbon. No graphitic layer has been observed on any diamond films. The surface amorphous carbon layer could be removed by oxygen plasma treatment after deposition and does not contribute to the superconductivity.

The superconducting transitions of the heavily boron-doped (111) and (100) diamond films are identified from the temperature dependence of magnetic susceptibility under zero-field cooling (ZFC) and field cooling (FC) using the SQUID as illustrated in fig. 2. Tc is determined to be 7.4 and 3.2 K for the (111) and (100) diamond films, respectively, from the onset of the diamagnetic transition under ZFC measurement. These values are almost the same as those for the $T_{c(offset)}$ measured by resistivity. The magnetic moment signals of the two diamond films do not saturate even at 1.78 K. It might be due to the inhomogeneous grain connectivity of diamond films. The hysteresis between FC and ZFC susceptibilities corresponds to the trapped flux under the field cooling condition.

The temperature dependences of the resistivity are measured on the (001) and (111) diamond films. The resistance characteristics of the (111) and (100) diamond films under the zero field condition as a function of temperature are shown in fig. 3a and b, respectively. Insets show the magnified views of the resistance at around $T_{c(onset)}$, which is determined as the temperature at which the resistance starts to disperse from the normal-state resistance. $T_{c(offset)}$ is defined as the temperature of zero resistance. The resistance of the (111) diamond film starts to disperse from the normal-state resistance (B = 9 T) at around 11.4 K ($T_{c(onset)}$) and vanishes below 7.4 K ($T_{c(offset)}$). On the other hand, $T_{c(onset)}$ and $T_{c(offset)}$ of the (100) diamond film are 6.3 K and 3.2 K, respectively, which are twice as low as those of the (111) film. The field dependences on $T_{c(onset)}$ and $T_{c(offset)}$ of the heavily boron-doped (111) and (100) diamond films as functions of temperature are illustrated in fig. 4. According to the weak-coupling BCS theory, the upper critical field $H_{c2}(0)$ can be estimated using the Werthamer – Helfand - Hohenberg (WHH) formula [11] as $H_{c2}(0) = -0.69(dH_{c2}/dT)_{T=T_c}T_c$ and the $T_{c(onset)}$ curves in Fig. 4, which lead to values of $H_{c2}(0)$ of 8.7 and 4.7 T for the (111) and (100) diamond films, respectively. These values are 2 – 3 times higher than all other values ever reported for diamond superconductivities [1-3]. These large $H_{c2}(0)$ values are due to the high Tc of our diamond and are advantageous for device applications.

Figure 5 shows the experimental results of $T_{c(onset)}$ and $T_{c(offset)}$ for heavily boron-doped (111) and (100), and polycrystalline diamond, as well as those reported in ref. 1, 2 and 3, as functions of boron concentration. The $T_{c(onset)}$ and $T_{c(offset)}$ of (111) diamond films are rose with increasing the boron content without saturation up to approximately 8 x $10^{21}$ cm$^{-3}$ (4.5 %). However, the Tc of (100) films seems to saturate at 8.5 x $10^{21}$ cm$^{-3}$ with the lower $T_{c(offset)}$ of 3.2 K in (100) diamond. A slightly lower $T_{c(offset)}$ of 2.9 K is obtained in (100) diamond with a higher boron content (1.4 x $10^{22}$ cm$^{-3}$, 7.9 %), which is the highest boron content ever reported. Tc of the heavily boron-doped polycrystalline film deposited by means of the CVD method with {111} facets almost fits the tendency for (111) diamond [2]. On the other hand, Tc of the heavily boron-doped HPHT diamond reported by Ekimov et al [1] fits the tendency for (100) diamonds. The saturation tendency of Tc in (100) diamond agrees with the previous result obtained at lower boron doping levels (5 – 7 x $10^{20}$ cm$^{-3}$) for (100) diamond films as reported by Burstarret et al. [3].

One of the reasons for heavily boron-doped (111) diamonds have an advantage on superconductivity compared with (100) diamonds is the higher carrier concentrations in (111) diamond. The hole carrier concentrations of boron-doped (100) diamond films are not more than the boron concentration measured by means of SIMS. On the other hand, the carrier concentrations of the boron-doped (111) diamond films, exhibiting higher Tc, exceed the boron concentrations in the films. Consequently, most (111) diamond films have higher carrier densities even at the level of boron content equivalent to that of the (100) diamonds. For example, the carrier concentration of the (111) diamond showing $T_{c(offset)}$ of 7.4 K with the boron concentration of 8.4 x $10^{21}$ cm$^{-3}$, is 1.3 x $10^{22}$ cm$^{-3}$. Similar phenomena, namely, that the carrier concentration of heavily boron-doped diamond exceeds the boron concentrations in the films, have been reported previously [12, 13]. Bustarret et al. explained the phenomenon is because of the much heavier effective mass for the holes or additional conduction channels [13]. From the Raman shift measurement as shown in fig. 1, the 500 cm$^{-1}$ band of the heavily boron-doped (111) diamond shifts to a lower wavenumber indicating the high density of boron pairs along <111> directions [10]. The bond length of this boron pair of 0.194 nm is much longer than that of C-C bond (0.154 nm) in diamond. This type of defect might expand the lattice of diamond [14], resulting in an increase in the density of states in the valence band. Nakamura et al. have indicated that an increase in the density of states was identified in superconducting polycrystalline diamond (4.2 % boron content) by utilizing soft X-ray emission and absorption spectroscopy [15]. Consequently, the Fermi level is observed 1.3 eV below the valence band maximum.

From the viewpoint of crystal growth, defects, such as B-B pairs, are easily formed during (111) growth rather than during (100) growth. In the case of the growth of diamond with high concentrations of the boron source using TMB, the growth species might contain many B-C composites. For the growth of (100) diamond, the B-C species is deposited to produce the surface dimer rows and to expand the terrace as step growth [16]. This growth mode forms B-B pairs with equal probability along every <111> directions and expands the lattice homogeneously. This expansion requires a high total potential energy. On the other hand, for the growth of (111) diamond the growth mode of which requires three atoms on the two surface diamond atoms [17], B-B pairs are prohibited along with the growth direction of [111], as shown in fig. 6a, because B on substitutional site 1 cannot form a chemical bond with the B in site 2 of the growth species. However, when substitutional B is in site 3 in fig. 4b, B of the growth species can be deposited on site 4. As a result, the B-B pair can be formed along the $[\bar{1}11]$ direction. The $[11\bar{1}]$ and $[1\bar{1}1]$ directions form the B-B pair with the same probability as in fig 4c and d. Accordingly, lattice expansions along the $[\bar{1}11]$, $[11\bar{1}]$ and $[1\bar{1}1]$ directions occur due to the anisotropic formation of B-B pairs. The total potential energy of (111) diamond with this kind of anisotropic strain might be lower than that of (100) diamond with homogeneous strain. In fact, (111) diamond is more stable with a high density of B-B pairs. Consequently, a higher density of B-B pairs can be formed compared with (100) diamond films with equivalent boron concentrations. Detailed experiments to determine the lattice expansion and density of states at the valence band are required to explain the large advantage of Tc of heavily boron-doped (111) diamond.

Recently, Tc of heavily boron-doped diamond has been estimated using a McMillan's equation [18] with calculated electron-phonon coupling coefficient $\lambda$ [4, 6]. Xiang et al. estimated Tc of 4.4 K at the boron content of 2.78% (4.9 x $10^{21}$ cm$^{-3}$), a value which agrees with our results for $T_{c(offset)}$ of (111) diamond. They also calculated the higher Tc of 23.6 K as a consequence of an increase in $\lambda$ when the boron content is increased to be 6.25 % (1.1 x $10^{22}$ cm$^{-3}$). In order to dope a higher density of boron to increase $\lambda$, a useful technique is co-doping using nitrogen. The co-doping technique is effective for increasing the solubility limit, and not only for reducing the ionization energy of impurities. Introducing a vacancy or a larger impurity atom such as substitutional Al, might be effective for expanding the lattice constant and hence for increasing the density of states at the valence band.

This work is supported in part by a Grant-in-Aid for Center of Excellence (COE) Research from the Ministry of Education, Culture, Sports, Science and Technology. This work is also supported in part by the Advanced Research Institute for Science and Engineering, Waseda University.

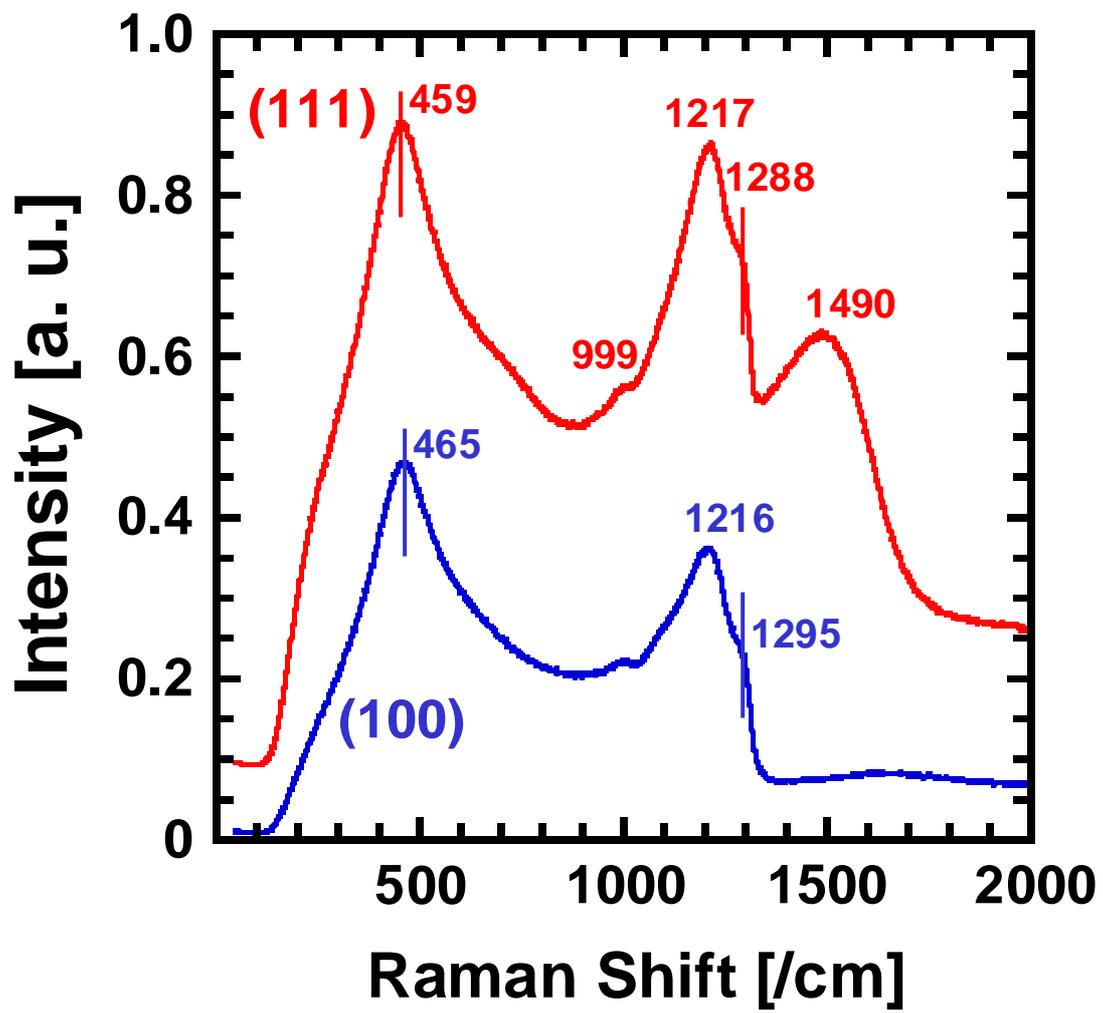

FIG. 1

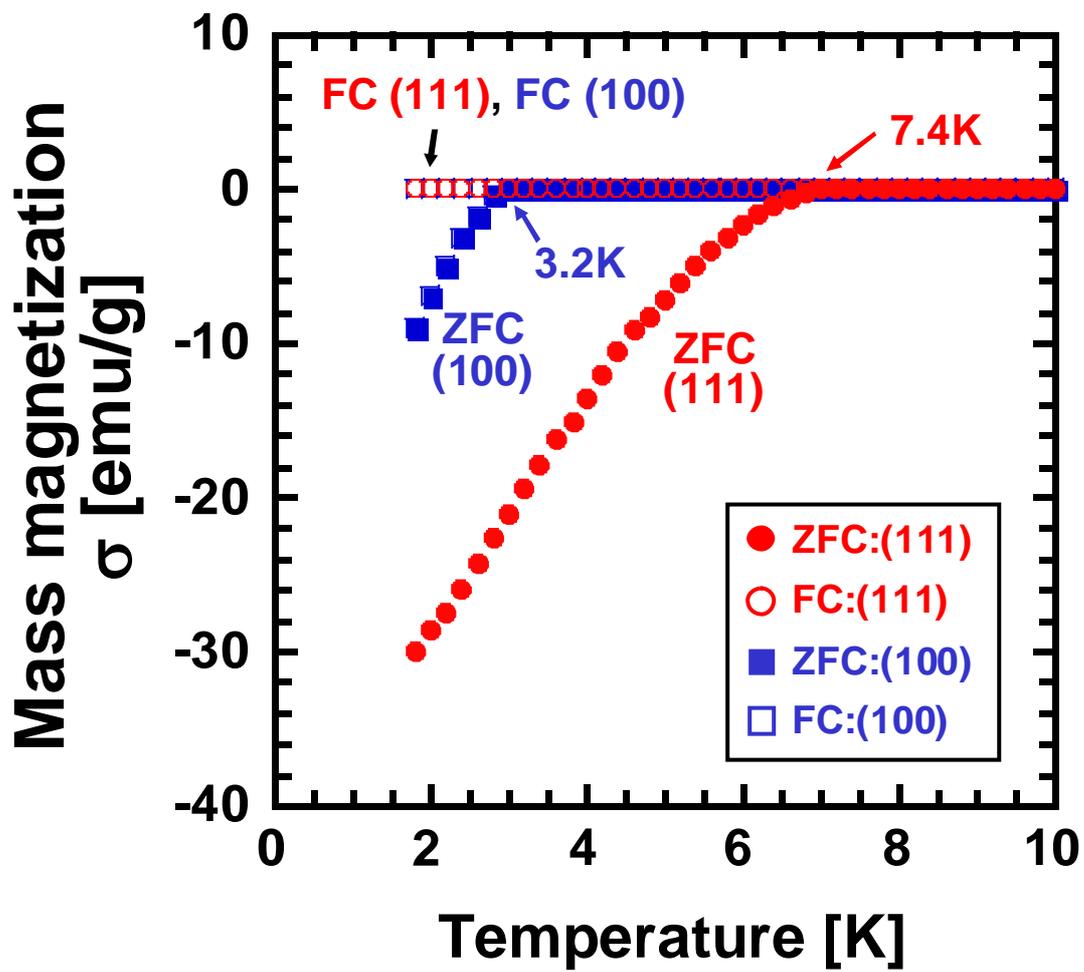

**FIG. 2**

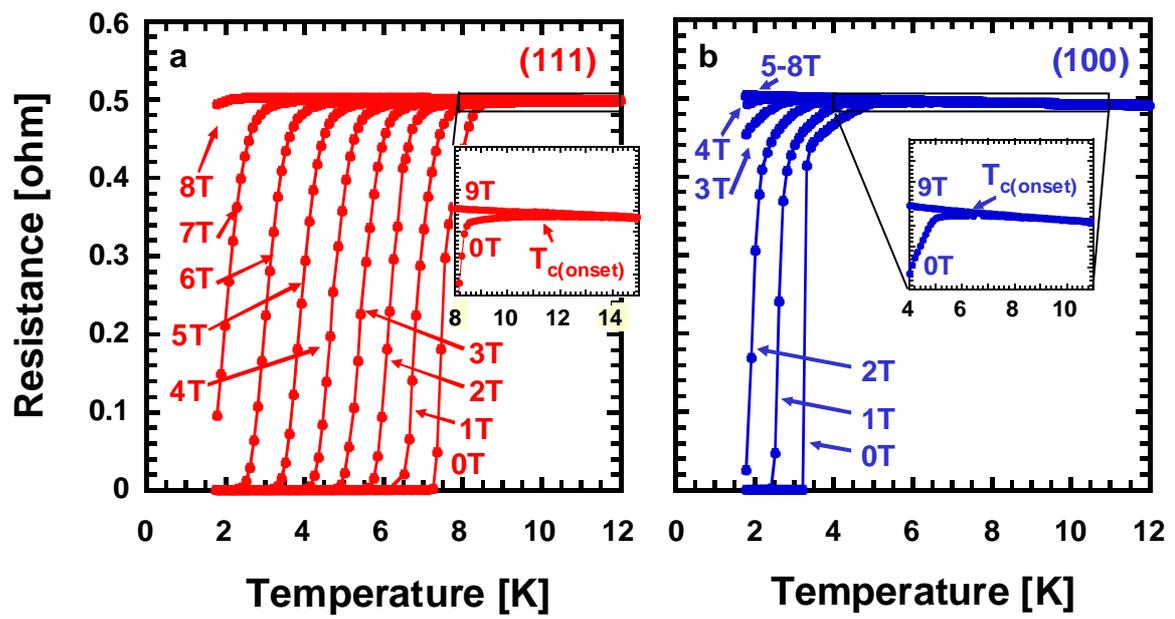

**FIG. 3**

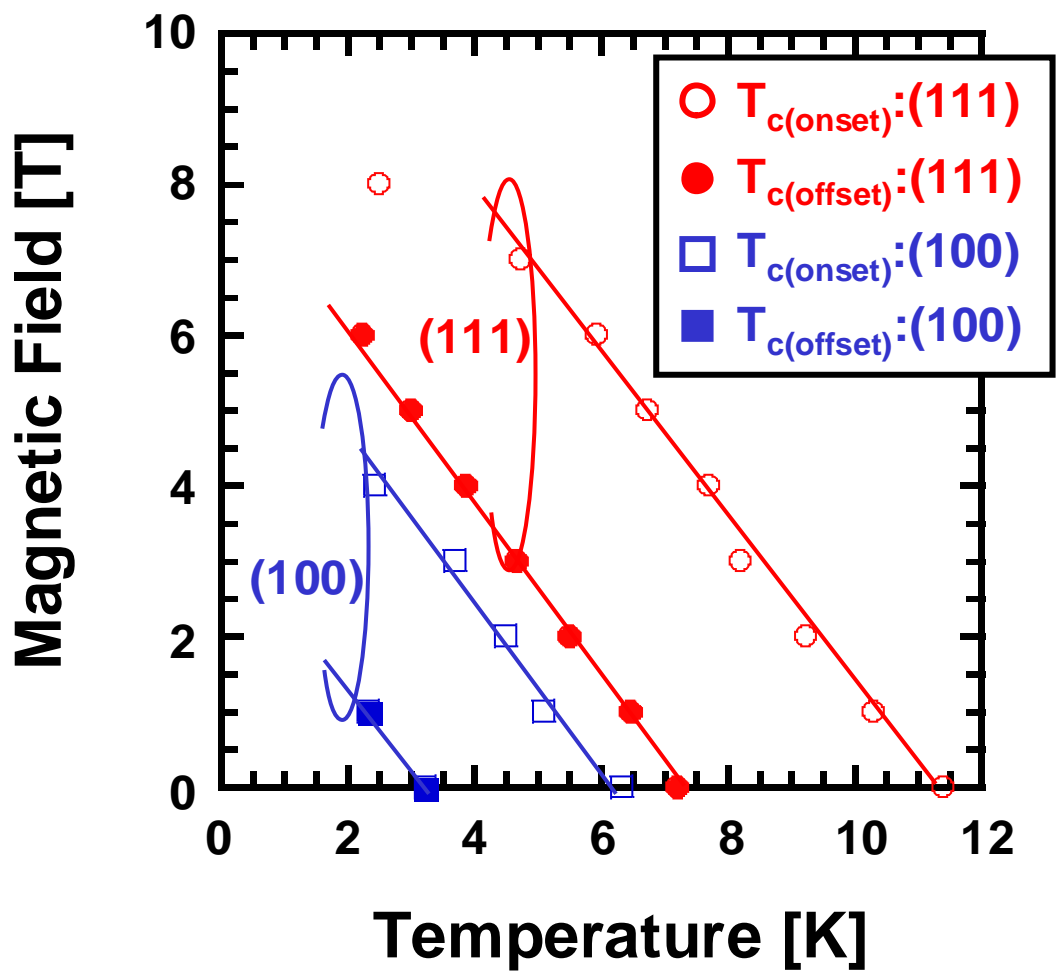

**FIG. 4**

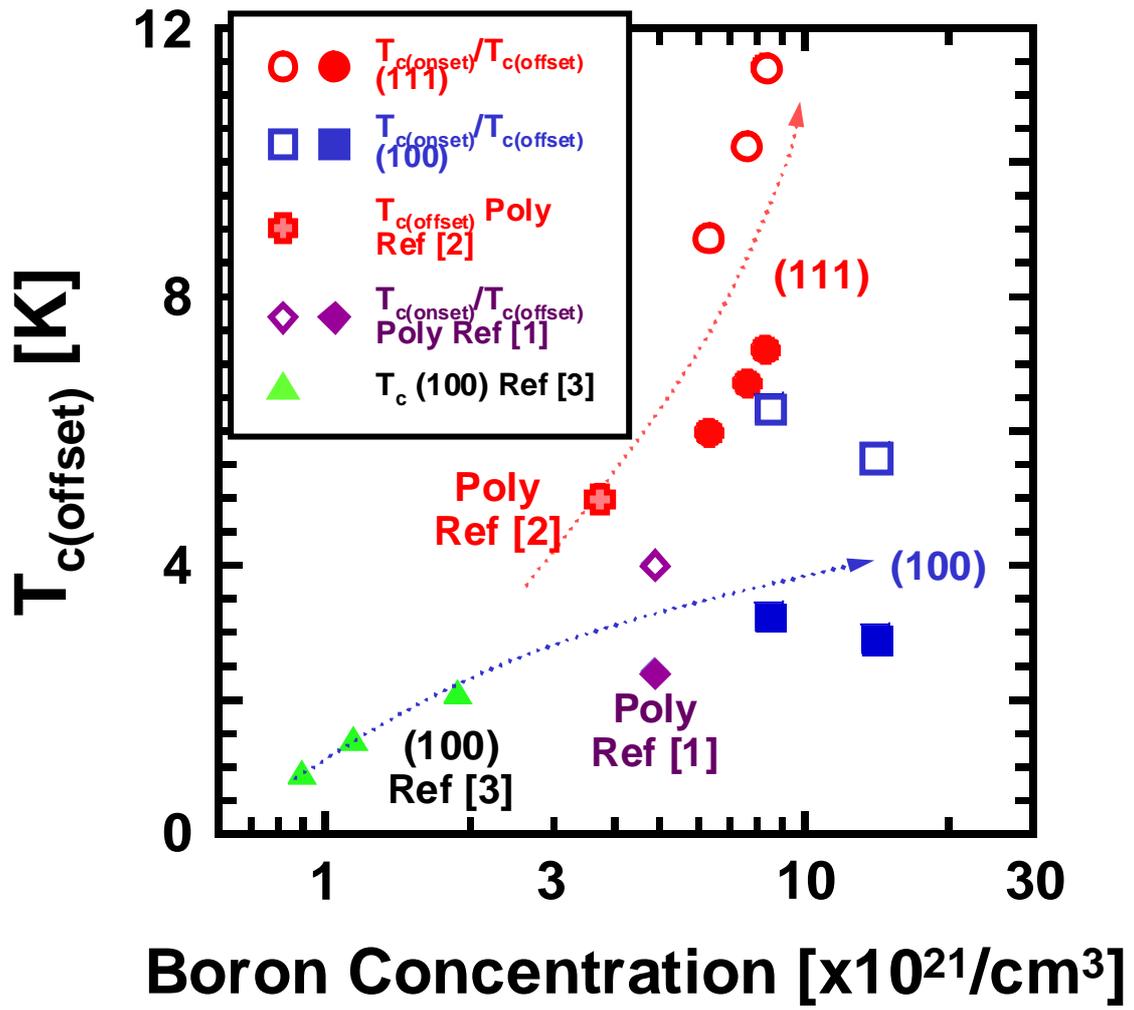

**FIG. 5**

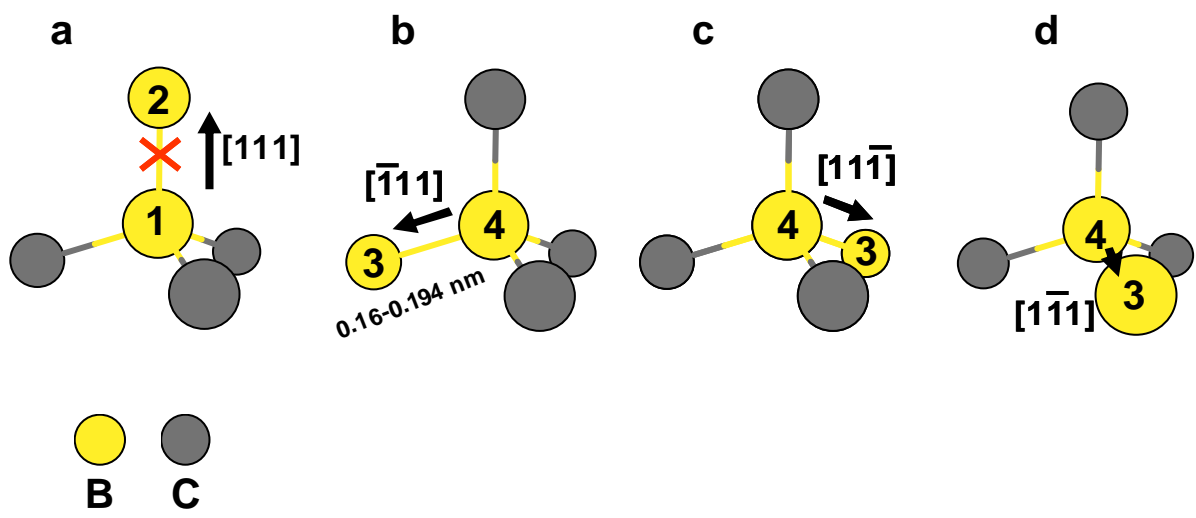

**FIG. 6**

Figure captions

Figure 1.

Raman shift of heavily boron-doped (111) and (100) diamond measured by microlaser Raman spectroscopy (632.8 nm). The 500 cm$^{-1}$ band on (111) diamond shifts to be lower wavenumbers than that of (100) diamond, which suggests that a more dense boron pairs exists in the (111) diamond film even at the equivalent boron concentrations of 8.4 x 10$^{21}$ cm$^{-3}$ for (111) and 8.5 x 10$^{21}$ cm$^{-3}$ for (100) diamond films, respectively.

Figure 2.

Mass magnetizations of heavily boron-doped (a) (111) and (b) (100) diamond films as functions of temperature. Data are shown for measurements under conditions of zero-field cooling (ZFC) and field cooling (FC) at 10 Oe. Tc of 7.4 and 3.2 K for (111) and (100) diamond films, respectively, are obtained from the onset of the diamagnetic transition.

Figure 3.

Electrical resistivity curves of homoepitaxially grown heavily boron-doped (a) (111) and (b) (100) diamond films under magnetic fields of up to 8 T. The highest $T_{c(onset)}$ and $T_{c(offset)}$ of 11.4 and 7.4 K, respectively, are obtained in (111) diamond film with the doping concentration of 8.4 x $10^{21}$ $cm^{-3}$.

Figure 4.

The field dependences of $T_{c(onset)}$ and $T_{c(offset)}$. The upper critical field $H_{c2}(0)$ of the (111) and (100) diamond films are estimated to be 8.7 and 4.7 T, respectively.

Figure 5.

Experimental results of $T_{c(onset)}$ and $T_{c(offset)}$ for polycrystalline, homoepitaxial (100) and (111) diamond films as a function of boron concentration. Tc values of (111) diamond films are more than twice as high as those of (100) films at the equivalent boron concentration.

Figure 6.

The schematic of the growth mechanism of the (111) diamond surface with a B-B pair. (a) The B-B pair along the [111] direction is prohibited due to the lack of a chemical bond on B (1). (b) A B-B pair can be formed along on the [$\bar{1}$11] direction between B(3) and B(4). The long bond length of the B-B pair expands the lattice of the diamond film. The (c) [11$\bar{1}$] and (d) [1$\bar{1}$1] directions form the B-B pair with the same probability as [$\bar{1}$11].

# References

1. E. A. Ekimov, V. A. Sidorov, E. D. Bauer, N. N. Melnik, N. J. Curro, J. D. Thompson, S. M. Stishov, *Nature* **428,** 542 (2004).

2. Y. Takano, M. Nagao, I. Sakaguchi, M. Tachiki, T. Hatano, K. Kobayashi, H. Umezawa, H. Kawarada, *Appl. Phys. Lett.* **85,** 2851 (2004).

3. E. Bustarret, J. Kacmarcik, C. Marcenat, E. Gheeraert, C. Cytermann, J. Marcus, T. Klein, *Phys. Rev. Lett.* **93,** 237005 (2004).

4. L. Boeri, J. Kortus, O. K. Andersen, *Phys. Rev. Lett.* **93,** 237002 (2004).

5. K. W. Lee, W. E. Picket, *Phys. Rev. Lett.* **93,** 237003 (2004).

6. H. J. Xiang, Z. Li, J. Yang, J. G. Hou, Q. Zhu, cond-mat 0406446.

7. G. Baskaran, cond-mat 0404286 (2004).

8. K. Ushizawa, K. Watanabe, T. Ando, I. Sakaguchi, M. Nishitani-Gamo, Y. Sato, H. Kanda, *Diam. Relat. Mater.* **7,** 1719 (1998).

9. P. Gonon, E. Gheeraert, A. Deneuville, F. Fontaine, L. Abello, G. Lucazeau, *J. Appl. Phys.* **78,** 7059 (1995).

10. M. Bernard, C. Baron, A. Deneuville, *Diam. Relat. Mater.* **13,** 896 (2004).

11. N. R. Werthamer, E. Helfand, P. C. Hohenberg, *Phys. Rev.* **147,** 295 (1966).

12. K. Nishimura, K. Das, J. T. Glass, *J. Appl. Phys.* **69,** 3142 (1991).

13. E. Bustarret, E. Gheeraert, K. Watanabe, *Phys. Stat. Sol. a,* **199,** 9 (2003).

14. F. Brunet, P. Germi, M. Pernet, A. Deneuville, E. Gheeraert, F. Laugier, M. Burdin, G. Rolland, *Diam. Relat. Mater.* **7,** 869 (1998).

15. J. Nakamura, *cond-mat* 0410144v3 (2004).

16. M. Tsuda, M. Hata, S. Oikawa, *Appl. Surf. Sci.* **107,** 116 (1996).

17  C. C. Battaile, D. J. Srolovitz, J. E. Butler, *Diam. Relat. Mater.* **6,** 1198 (1997).

18  W. L. McMillan, *Phys. Lev.* **167,** 331 (1968).